\documentstyle[12pt]{article}
\def\beq{\begin{equation}}
\def\eeq{\end{equation}}

\begin{document}
\begin{titlepage}
\begin{flushright}
BUTP-94/2 \\
CERN-TH.7215/94 \\
\end{flushright}
\vspace{0.2in}
\begin{center}
{\Large \bf Next-to-leading-order temperature corrections \\
to correlators in QCD} \\
\vspace{0.4in}
{\bf V.L. Eletsky$^{\dagger}$,} \\
{\em Institute for Theoretical Physics, Berne University, \\
Sidlerstrasse 5, CH-3012 Berne, Switzerland  \\
and \\
CERN, Geneva, Switzerland} \\
\vspace{0.2in}
{\bf and } \\
\vspace{0.2in}
{\bf B.L. Ioffe} \\
{\em Institute of Theoretical and
Experimental Physics, Moscow 117259, Russia} \\
\vspace{0.5in}
{\bf   Abstract  \\ }
\end{center}

Corrections of order $T^4$ to vector and axial current correlators in QCD
at a finite temperature $T<T_c$ are obtained using dispersion relations for the
amplitudes of deep inelastic scattering on pions.
Their relation with the operator product expansion is presented.
An interpretation of the results in terms of $T$-dependent meson masses is
given: masses of $\rho$ and $a_1$ start to move with temperature in
order $T^4$.

\vskip1.5in
\begin{flushleft}
CERN-TH.7215/94 \\
March 1994
\end{flushleft}
\hrule height .2pt width 3in
\noindent$^{\dagger}$On leave of absence from the Institute of
Theoretical and Experimental Physics, Moscow 117259, Russia.
\end{titlepage}

In the last years there has been an increasing interest
in the study of the
current correlators in QCD at finite temperatures. The hope is that
investigating the same correlators, both at high temperatures, where the
state of quark-gluon plasma is expected,
and at low temperatures, where the
hadronic phase persists, a clear signal for a phase transition could be
found. For a review of calculations of correlators performed
by various analytical methods and in the lattice simulations,
see, e.g., Ref.~\cite{sh}.

%ins1
The study of temperature dependence of current
correlators is interesting in many aspects. At small distances the
correlators are expressed through the operator product expansion (OPE) in
terms of matrix elements of the operators of low dimension. In this way, the
temperature dependence of these matrix elements manifests itself in the
temperature dependence of the correlators and visa verse.  At $T=0$ using
dispersion relations the correlators can be expressed in terms of
contributions of hadronic states.  Then, using some theoretical tools
(differentiation, Borel transformation, etc) it is possible to enhance the
contribution of lowest hadronic states. Therefore, knowledge of
$T$-dependence of hadronic correlators can give us an information about the
$T$-dependence of masses of the lowest hadronic states.
Since in the approach of QCD sum rules, these masses are
determined by the matrix elements of operators in OPE,
the $T$-dependences of both are interrelated.

As argued in Ref.~\cite{ei}, in carrying out this program
certain wrong steps had been taken,
and a misunderstanding existed in the literature. Thus, some authors
maintained that the thermal shift of hadron masses occurs already in
the lowest order in temperature, $O(T^2)$. However, the general
statement is that hadron masses do not move in this order~\cite{ei,dei}.
This statement is very general:
it is based only on PCAC and current algebra.
(The result that the nucleon pole
does not move in order $T^2$ was obtained in the chiral
perturbation theory in Ref.~\cite{ls} and by considering a
current correlator in Ref.~\cite{e2}).
The only interesting physical phenomenon that occurs
in this order is the parity mixing, i.e. the appearance
of states with opposite parity in a given channel and, in some cases,
also an isospin mixing. The latter arises in baryon current correlators
where, for example, in the current with the
quantum numbers of $\Lambda$ there
appears a $\Sigma$ pole, and in the nucleon channel there appear poles
corresponding to baryon resonances with $J^{P}=\frac{1}{2} ^{\pm}$ and
$T=\frac{3}{2} ,\;\frac{1}{2}$. For the case of vector
and axial currents in the chiral limit, this mixing is given by

\begin{eqnarray}
C_{\mu\nu}^{V}(q,T)&=&(1-\epsilon)C_{\mu\nu}^{V}(q,0)+
\epsilon C_{\mu\nu}^{A}(q,0) \nonumber\\
C_{\mu\nu}^{A}(q,T)&=&(1-\epsilon)C_{\mu\nu}^{A}(q,0)+
\epsilon C_{\mu\nu}^{V}(q,0)
\label{mix}
\end{eqnarray}
where $C_{\mu\nu}^{V(A)}(q,T)$ are the correlators of
$V$ and $A$ currents at finite temperature,
$C_{\mu\nu}^{V(A)}(q,0)$ are the same correlators at $T=0$,

\begin{equation}
\epsilon =\frac{T^2}{6F_{\pi}^2},
\label{eps}
\end{equation}
and $F_{\pi}=93$\, MeV is the pion decay constant.

%ins2
At $T\neq 0$ both $C_{\mu\nu}^V$ and $C_{\mu\nu}^A$ have transverse
and longitudinal parts

\begin{equation}
C_{\mu\nu}^{V(A)}(q,T)=
(-g_{\mu\nu}q^2 +q_{\mu}q_{\nu})C_{t}^{V(A)}(q^2,T)
+q_{\mu}q_{\nu}C_{l}^{V(A)}(q^2,T)
\label{tl}
\end{equation}
At $T=0$ $C_{l}^{V}(q^2,0)=0$, but $C_{l}^{A}(q^2,0)$ is non-zero
and in the chiral limit (when light quarks and pions are massless)
is given by the one-pion contribution

\begin{equation}
C_{l}^{A}(q^2,0)=\frac{F_{\pi}^2}{q^2}
\label{l}
\end{equation}
According to Eq.~(\ref{mix}), at $T\neq 0$ the longitudinal
part (the pion pole) appears also in the vector channel.

If $C^{V,A}(q,0)$ are represented through dispersion relations
by contributions of the physical states in the $V$ and $A$ channels
($\rho$, $a_1$), then according to Eq.~(\ref{mix}) the poles
that are in the r.h.s. of Eq.~(\ref{mix}), i.e. at $T=0$, appear at the
same positions in the l.h.s. Therefore, in order $T^2$ the poles
corresponding to $\rho$ and $a_1$ do not move~\cite{ei}.
An important consequence of Eq.~(\ref{mix}) is that at $T\neq 0$ in the
vector (transverse) channel apart from the poles corresponding to vector
particles, there arise poles corresponding to axial particles and vice
versa, i.e. a sort of parity mixing phenomenon occurs.
The manifestation of this phenomenon is in complete
accord with the general considerations presented
in Ref.~\cite{ei}:
the appearance of the $\rho$ ($a_1$) pole in the axial (vector) channel
corresponds to singularities in the $s$-channel. In the same way a pion
pole appears in the longitudinal part of the vector channel.

In this paper we show that in the next order, $O(T^4)$,
such a simple picture, where the current
correlator at finite temperature is represented by the superposition of
$T=0$ correlators, does not take place. Interpreted in terms of
temperature-dependent poles, it would mean that masses are shifted in
this order.
%ins3
In what follows we consider only the transverse part of the correlator
because the $T$-dependence of the pion mass and the decay constant
$F_{\pi}$ was thoroughly investigated earlier~\cite{leut}, and we can
say nothing new here.

The thermal correlation functions in Eq.~(\ref{mix}) are defined as

\begin{equation}
C_{\mu\nu}(q,T)=
\langle i\int d^4 x e^{iqx}T\{j_{\mu}(x),j_{\nu}(0)\}\rangle _T =
\frac{\sum_{n}\langle n|i\int d^4 x e^{iqx}T\{j_{\mu}(x),j_{\nu}(0)\}
e^{-H/T}|n\rangle}{\sum_{n}\langle n|e^{-H/T}|n\rangle},
\label{c}
\end{equation}
where $H$ is the QCD Hamiltonian, the sum is over all states of
the spectrum, and $j_{\mu}(x)$ is either a vector or an axial current,
$j_{\mu}(x)=
(1/2)(\bar{u}\gamma_{\mu}(\gamma_{5})u - \bar{d}\gamma_{\mu}(\gamma_{5})d)$.
It is assumed that $q^2$ is space-like, $Q^2=-q^2>0$, and $Q^2$
is much larger than a characteristic hadronic scale,
$Q^2\gg R_{c}^{-2}$,
where $R_c$ is the confinement radius, $R_{c}^{-1}\sim 0.5\:$GeV.

We consider the case of temperatures $T$ below the phase
transition temperature $T_c$. In principle, the summation over $n$ in
Eq.~(\ref{c}) can be performed over any complete set of states
$|n\rangle$ in the
Hilbert space. It is clear however, that at $T<T_c$ the suitable set of
states is the set of hadronic states, but not the quark-gluon basis.
Indeed, in this case the original particles forming
the heat bath, which is probed by
external currents, are hadrons. The summation over the quark-gluon
basis of states would require to take into account
the full range of their interaction.
In connection with consideration of current correlators
at finite $T$, this
point was first explicitly made in Ref.~\cite{dei}. In the early
papers~\cite{prev} devoted to the extension of QCD sum rules to
finite temperatures the summation over $|n\rangle$ at low $T$ was
performed in the quark-gluon basis without account of confinement. In a
recent paper~\cite{hkl} QCD sum rules at finite $T$ were rewritten using
the pion basis.

At $T$ well below the phase transition temperature $T_c$ an expansion in
$T$ can be performed.  The main contribution comes from the pion states,
$|n\rangle =|\pi\rangle ,\; |2\pi\rangle ,\;...$
%ins4
In this paper we restrict ourselves to the chiral limit, when $u$, $d$
quarks and pions are massless. The corrections to the chiral limit will
be considered in a separate publication.

In the chiral limit there are two parameters in the low $T$ expansion.
One parameter appears when the pion momenta, $p\sim T$, in the matrix
elements in Eq.~(\ref{c}) can be neglected. Then, the matrix elements
in Eq.~(\ref{c}) can be calculated using PCAC and current algebra.
The powers of $T^2$ arise due to phase space integration with the
Bose factor. In this case the expansion parameter is $T^2/F_{\pi}^2$:
the one-pion contribution is proportional to $T^2/F_{\pi}^2$, the
two-pion contribution is of order $T^4/F_{\pi}^4$, etc~\cite{dei}.
In the order $T^2$ there are only terms of this type. It is clear
that in any order $(T^2)^n$ the terms of this type are expressed
through the vector and axial vacuum correlators
$C_{\mu\nu}^{V(A)}(q^2,0)$ and, as a consequence, do not result in thermal
shifts of hadron masses.
The other expansion parameter appears if non-vanishing pion momenta
in the matrix elements in Eq.~(\ref{c}) are taken into account.
Since the characteristic distances in Eq.~(\ref{c}) are of order
$x^2\sim 1/Q^2$, the expansion parameter in this case is $T^2/Q^2$.
It is also worth to mention that the contributions of massive hadronic
states $|n\rangle$ are exponentially suppressed as $\exp(-m_n /T)$.

Let us start with the calculation of $T^4$ terms of the first type.
The matrix elements over two-pion
states in the limit $p\to 0$ give a $T^4$ contribution. But this is not
just a second iteration of the procedure used to obtain the one-pion
matrix element. It is also necessary to take into account the interaction
between the pions in the initial and the finite states. This can be
illustrated by the example of $T^4$ terms in the $T$-dependence of the quark
condensate\cite{leut}:

\begin{equation}
\langle\bar{q}q\rangle _T = \langle\bar{q}q\rangle _0
\left( 1-\frac{3}{4}\epsilon -\frac{3}{32}\epsilon^2\right)
\label{qq}
\end{equation}
The
$\epsilon$ and $\epsilon^2$ terms here come from one- and two-pion matrix
elements, respectively. However, not accounting for the initial (finite)
state interaction of pions in the two-pion state would give
$9\epsilon^2 /32$ instead.
The interaction amplitude for $\pi^b \pi^d \to \pi^a \pi^c$ to
zeroth order in pion momenta is given by

\begin{equation}
\frac{m_{\pi}^2}{F_{\pi}^2}
(3\delta_{ac}\delta_{bd}-\delta_{ab}\delta_{cd}-\delta_{ad}\delta_{bc})\, .
\label{4pi}
\end{equation}
The $m_{\pi}^2$ in the numerator of the above amplitude cancels against
the pion mass in the pion propagator at $p\to 0$ and contributes
$-3\epsilon^2 /8$ to the $T$-dependence of $\langle\bar{q}q\rangle$,
making the correct total of $-3\epsilon^2 /32$.

Similarly, it is easy to show that in the case of vector and axial
correlators the two-pion matrix elements with the account of
initial (finite) state interaction of pions result
in the following expressions for the transverse part of $V$ and $A$
correlators:

\begin{equation}
C_{\mu\nu}^{V(A)}(q,T)=\left(-g_{\mu\nu}q^2 +q_{\mu}q_{\nu}\right)
\left( C_{V(A)}(q^2)+\left(\epsilon-\frac{1}{2}\epsilon^2\right)
\left( C_{A(V)}(q^2)-C_{V(A)}(q^2)\right)\right)\, .
\label{mix1}
\end{equation}
As was expected, the $T^4/F_{\pi}^4$ terms above are again
expressed through
correlators at $T=0$, and they do not result in a thermal mass shift.
Also, these terms are Lorentz invariant, since only the tensor
$-g_{\mu\nu}q^2+q_{\mu}q_{\nu}$ appears here.

Consider now the terms of the second type, that arise from
non-zero pion momenta. It is well known from the description of
deep inelastic electron-hadron scattering \footnote{We are now considering
the case of a vector correlator. However, the axial correlator in the chiral
limit has the same tensor structure, and the results obtained in the vector
case may be directly applied to the axial one.} that a matrix element of the
product of two vector currents may be represented using two tensor
structures

\begin{eqnarray}
\lefteqn{T_{\mu\nu}^{\pi}(p,q)=i\int d^4 x e^{iqx}
\langle\pi (p)|T\{j_{\mu}(x),j_{\nu}(0)\}|\pi (p)\rangle } \nonumber\\
& & =\left( -g_{\mu\nu}+\frac{q_{\mu}q_{\nu}}{q^2}\right)
T_{1}(\nu ,q^2)+
\left( p_{\mu}-\nu\frac{q_{\mu}}{q^2}\right)
\left( p_{\nu}-\nu\frac{q_{\nu}}{q^2}\right)T_{2}(\nu ,q^2)\, ,
\label{dis}
\end{eqnarray}
where $\nu =pq$, $T_1$ is dimensionless, and $T_2$ has dimension
$mass^{-2}$.
The corresponding contribution to the thermal correlation function is
obtained by integrating the above equation over the pion phase space
with the Bose factor. And it is the second term in the r.h.s. of
Eq.~(\ref{dis}) that, after this integration, provides the expected Lorentz
non-invariant contribution to the thermal correlator. In terms of the OPE,
the function $T_2$ is contributed only by averages of Lorentz non-scalar
operators, while $T_1$ receives contributions from both Lorentz scalar
and non-scalar operators\cite{eek}.

It is well known\cite{book} that the function $T_2$ satisfies a
dispersion relation without subtractions:

\begin{equation}
T_2 (\nu ,q^2)=\frac{2}{\pi}\int_{Q^2/2}^{\infty}
\frac{\nu^{\prime}{\rm Im}\, T_2 (\nu^{\prime},q^2)d\nu^{\prime}}
{{\nu^{\prime}}^2 -\nu^2}\, .
\label{disp2}
\end{equation}
Having in mind the subsequent integration over $p$, we are interested
in $T_2 (0,q^2)$. Then, using the relation between the imaginary part
of $T_2$ and the structure function $F_2$

\begin{equation}
{\rm Im}\, T_2 (\nu ,q^2)=\frac{2\pi}{\nu}\, F_2 (x,q^2)\, ,
\label{Im2}
\end{equation}
where $x=Q^2/2\nu$ is the standard deep inelastic scaling variable, we get

\begin{equation}
T_2 (0,q^2)=\frac{8}{Q^2}\,\int_{0}^{1} F_{2}(x,q^2)dx\, .
\label{T2}
\end{equation}
Similarly, the function $T_1 (\nu, q^2)$ satisfies a dispersion relation
with one subtraction,

\begin{equation}
T_1 (\nu, q^2)-T_1 (0,q^2)=\frac{2\nu^2}{\pi}\,
\int_{Q^2/2}^{\infty}
\frac{{\rm Im}\, T_{1}(\nu^{\prime},q^2)d\nu^{\prime}}
{\nu^{\prime}({\nu^{\prime}}^2 -\nu^2)}\, .
\label{disp1}
\end{equation}
Again, using the relation between ${\rm Im}\, T_1$ and the structure function
$F_1$

\begin{equation}
{\rm Im}\, T_1 (x,q^2) =2\pi F_1 (x,q^2),
\label{Im1}
\end{equation}
we obtain

\begin{equation}
T_1 (\nu, q^2)-T_1 (0,q^2)=\frac{8\nu^2}{Q^4}\,\int_{0}^{1}
2xF_{1}(x,q^2)dx\, .
\label{T1}
\end{equation}
%ins5
The subtraction constant $T_{1}(0,q^2)$ corresponds to zero momenta
of the initial and final pions and was already accounted for in the terms
proportional to $T^2/F_{\pi}^2$. It is worth to mention that at this stage
the assumption $Q^2>>R_{c}^{2}$ was not used. In the derivation of Eqs.
(\ref{T2}) and (\ref{T1}) it was assumed that $\nu <<Q^2$, which is
equivalent in the chiral limit to $Q>T$. The same assumption was sufficient
for the derivation of Eq.~(\ref{mix1}). Therefore, Eqs.~(\ref{mix1}),
(\ref{T2}) and (\ref{T1}) are correct also at moderate $Q^2\sim 1\,$GeV$^2$.

At higher $Q^2$, in the scaling region, the integrals in Eqs.~(\ref{T2})
and (\ref{T1}) are equal, since in this region
$2xF_{1}(x,Q^2)=F_{2}(x,Q^2)$. The integral

\begin{equation}
M_2 =\int_{0}^{1} F_{2}(x)dx
\label{M2}
\end{equation}
is the second moment of the structure function and in the parton model has
the meaning of the fraction of the pion momentum carried by quarks. In our
normalization of currents

\begin{equation}
M_2 =\frac{1}{4}\cdot\frac{1}{3}\int_{0}^{1}x\, dx\,
\sum_{q,a} [q_{\pi ^a}(x)+\bar{q}_{\pi ^a}(x)]=
\frac{1}{2}\int_{0}^{1}x\, dx\, [v(x)+2s(x)],
\label{dist}
\end{equation}
where $v(x)$ and $s(x)$ are
the distributions of valence and sea quarks in the pion (see, e.g.
\cite{book}). The factor $1/4$ in Eq.~(\ref{dist}) comes from the definition
of the currents in Eq.~(\ref{c}), and the factor $1/3$ arises due to
averaging over $\pi^{+},~\pi^{-},~\pi^{0}$ in the heat bath.

The distributions $v(x)$ and $s(x)$ were
parametrized in Ref.~\cite{exp} to fit the experimental data on
the Drell-Yan process $\pi + N \to l^{+}l^{-} + X $ and on the direct
photon production $\pi + N \to \gamma + X$, and it was found that
$M_2\approx 0.12$, which is somewhat lower than the result $0.15\pm 0.02$ of
Ref.~\cite{bb}, where this quantity was obtained from QCD sum rules for a
correlation function in external symmetric tensor field, and close to
$0.11\pm 0.03$ as estimated in Ref.~\cite{eek}.  These numbers correspond to
the normalization point $\mu =1$\, GeV.

Now, to obtain the corresponding contribution to the thermal
correlator, one has just to do the Bose-weighted integrals
over the pion momentum and sum over the three pions

\begin{equation}
3\int\frac{d^3 p}{(2\pi)^3 2p}\,\frac{1}{\exp(|up|/T)-1}\,
T_{\mu\nu}^{\pi}(p,q)
\label{int}
\end{equation}
($u$ is the 4-velocity of the heat bath), which gives together
with Eq.~(\ref{mix1})

\begin{equation}
C_{\mu\nu}(q,T)=
\left( -g_{\mu\nu}q^2 +q_{\mu}q_{\nu}\right)
C_{1}(T,q^2)+u^{t}_{\mu}u^{t}_{\nu}C_{2}(T,q^2)\, ,
\label{cmn}
\end{equation}
where $u^{t}_{\mu}=u_{\mu}-(uq)q_{\mu}/q^2$,

\begin{eqnarray}
C_{1}(T,q^2)&=&C_{V}(0,q^2)+\left(\epsilon -\frac{1}{2}\epsilon^2\right)
\left( C_{A}(q^2)-C_{V}(q^2)\right) +
\frac{1}{2}\, c\,\frac{1+2{\bf q}^2/q^2}{q^4}\, T^4 \nonumber\\
C_{2}(T,q^2)&=&-c\,\frac{T^4}{q^2}\, ,
\label{c12}
\end{eqnarray}
and

\begin{equation}
c= \frac{8\pi^2 M_2}{15}\, .
\label{coef}
\end{equation}
The low temperature expansion of $C_1$ contains powers of both $T^2/F_{\pi}^2$
and $T^2/Q^2$, but powers of $T^2/F_{\pi}^2$ are absent in $C_2$.
Notice, that while all three pion charge
states contribute to $T^2/Q^2$ terms, only two of them contribute to
$T^2/F_{\pi}^2$ terms.
The same formulae hold for the axial correlator, with the obvious change
$V\leftrightarrow A$.

The above formulae for the thermal correlator may be considered also
from the viewpoint of the OPE for the
correlation function(see, Refs.~\cite{hkl,eek}).
The OPE itself of course carries no information                       hey
about the state over which the matrix element of the operators is
considered, or about the heat bath, in case of finite $T$. It contains
operators of arbitrary Lorentz spin $s$ and twist $t$. When vacuum
correlators are considered, only scalar, $s=0$, operators contribute,
while non-scalar, $s\neq 0$, operators drop out in the averaging.
When averaging over a hadron state or over a heat bath, the $s\neq 0$
operators do contribute, since there is an additional vector in the
problem, the target momentum $p$ or the 4-velocity of the heat bath $u$.
The matrix elements of $s=0$ operators over pions or the heat bath may be
estimated (if the pion momenta can be neglected) using PCAC, which relates
them to vacuum averages. The $s=0$ operators do not give $(T^2/Q^2)^n$
terms and therefore do not result in thermal mass shifts.
The general expression for the hadron matrix element of an $s=n$,
$n\ge 2$ operator is $\langle p|\hat{O}_{\mu_1 ,\mu_2 ,...\mu_n}|p\rangle =
ap_{\mu_1}p_{\mu_2}...p_{\mu_n}$ (in the chiral limit there are no trace
terms $\sim g_{\mu_i\mu_j}$). These matrix elements cannot be reduced by
PCAC to vacuum averages and are new non-perturbative parameters. Their
Bose-weighted integrals over the pion momenta are
$T$-dependent $s\neq 0$ condensates which are suppressed as $T^s$ compared
to the $T$-dependent parts of $s=0$ condensates.

It is clear that in terms of the OPE the function $C_1$
is contributed both by $s=0$ and $s\neq 0$ condensates,
as is $T_1$. However, $C_2$ and $T_2$ are related only to the
$s\neq 0$ condensates.

In the chiral limit a difference in the
$s=0$ operators for the vector and axial correlators appears on the level of
4-quark operators
\footnote{The gluon condensate gets its $T$ dependence
only in order $T^8$~\cite{dob}, anyway.}.
A good consistency check of the calculation of
correlators in the pion gas approximation is on whether the
$T$ dependences of the $s=0$ 4-quark condensates in $C_1^{V(A)}$ match
the $V-A$ mixing in the first of Eqs.~(\ref{c12}). This indeed
turned out to be the case, as demonstrated (to order $T^2$) in
Ref.~\cite{e4}, and also in Ref.~\cite{k} for baryonic currents.
The correlation in $T$-dependences of the $s=0$ condensates in opposite
parity channels is not accidental and is related to the {\em scattering} of
thermal pions on the currents. Notice also, that since

\begin{equation}
C_1^V (T)-C_1^A (T)=(1-2\epsilon+\epsilon^2)\,
(C_1^V (0)-C_1^A (0))
\label{w}
\end{equation}
this correlation exactly satisfies Weinberg sum rules
generalized~\cite{ks} to finite $T$.

Among the non-singlet condensates, the leading contribution to $C_2$
at low $T$ comes from the lowest spin, $s=2$, which corresponds to the
$T^4$ behavior. In the leading twist there are two $s=2$ operators which
are related to the energy-momentum tensors of quarks,
$\theta_{\lambda\sigma}^q$, and gluons, $\theta_{\lambda\sigma}^G$,

\begin{eqnarray}
\theta_{\lambda\sigma}^q &=&\frac{i}{2}
(\bar{q}\gamma_{\lambda}D_{\sigma}q+\bar{q}\gamma_{\sigma}D_{\lambda}q),
{}~~~q=u\, ,\, d
\nonumber \\
\theta_{\lambda\sigma}^G &=&G_{\lambda\alpha}^{a}G_{\alpha\sigma}^{a}-
\frac{1}{4}g_{\lambda\sigma}G_{\alpha\beta}^{a}G_{\beta\alpha}^{a}\, .
\label{theta}
\end{eqnarray}

%ins6
Explicit expressions for the contribution of these operators to $T_1$ and
$T_2$ can be obtained from the general formulae of the theory of deep
inelastic scattering (see, e.g.~\cite{b}). We present here the result for
the case, when the longitudinal structure function
$F_L =2xF_{1}(x)-F_{2}(x)$ is neglected and only QCD corrections
proportional to $\alpha_{s}\ln (Q^2/\mu^2)$ are retained. It can be shown
that since all pion charge states are equally populated in the heat bath,
only flavour-singlet operators contribute to the structure functions. Then

\begin{eqnarray}
\lefteqn{T_{1}(\nu, q^2)=\frac{\nu}{2x}\, T_{2}(\nu ,q^2)} \nonumber\\
& & =\,\frac{q_{\lambda}q_{\sigma}}{q^4}
\left[\left( 1-\frac{8}{9}\frac{\alpha_s}{\pi}\ln(Q^2/\mu^2)\right)
\langle\pi|\Sigma_{q}\theta_{\lambda\sigma}^{q}|\pi\rangle +
\frac{\alpha_s}{8\pi}\ln(Q^2/\mu^2)
\langle\pi|\theta_{\lambda\sigma}^{G}|\pi\rangle\right]\, ,
\label{ope}
\end{eqnarray}
where the averaging over the three pion charge states must be performed.
It is easy to see that the contribution of $\theta_{\lambda\sigma}^G$ to
Eq.~(\ref{ope}) is small at $Q^2\sim 1$~GeV$^2$.

The pion matrix element of the total energy-momentum tensor is just a
normalization constant,

\begin{equation}
\langle \pi (p)|\theta_{\mu\nu}^u +\theta_{\mu\nu}^d
+\theta_{\mu\nu}^G|\pi (p)\rangle =2p_{\mu}p_{\nu}
\label{tot}
\end{equation}
($\langle \pi (p)|\pi (p^{\prime})\rangle = (2\pi)^3\, 2E\,
\delta^{(3)}({\bf p}-{\bf p}^{\prime})$).
%ins7
On the other hand, we have

\begin{equation}
\langle \pi (p)|\theta_{\mu\nu}^u +\theta_{\mu\nu}^d |\pi (p)\rangle =
8M_2 p_{\mu}p_{\nu}\, .
\label{ud}
\end{equation}
So, if we define also a constant
$b$ as $\langle\pi| \theta_{\mu\nu}^G|\pi\rangle =bp_{\mu}p_{\nu}$, then
$8M_2 +b=2$.  The constant $b$ enters the matrix element
$\langle\pi |{\bf E}^2 +{\bf B}^2|\pi\rangle$ and also $\langle {\bf E}^2
+{\bf B}^2 \rangle _T$. It was determined in Ref.~\cite{eek} from a duality
type of consideration:  $b=1.16\pm 0.14$ at $\mu =1$ GeV. This is in accord
with the estimates for $M_2$ obtained in Refs.~\cite{hkl,bb} and with the
statement that gluons carry about 50% of the total momentum as in the case
of nucleons.

Now, we would like to discuss a possibility to interpret
the $T^4/Q^2$ corrections to the correlators in Eqs.~(\ref{c12})
in terms of particle thermal mass shifts.
It is convenient to use the standard representation~\cite{prev} of the vector
correlator in a medium in terms of two invariant functions $C_l$ and
$C_t$, which in the rest frame of the heat bath ($u=(1,{\bf 0})$)
are defined as

\begin{eqnarray}
C_{00}^{T}&=&{\bf q}^2 C_{l}^{T}  \nonumber\\
C_{ij}^{T}&=&(\delta_{ij}-\frac{q_i q_j}{{\bf q}^2})C_{t}^{T}+
\frac{q_i q_j}{{\bf q}^2}q_0^2 C_l^T\, ,
\label{lt}
\end{eqnarray}
or in terms of $C_1$ and $C_2$

\begin{eqnarray}
C_t &=&q^2 C_1 \nonumber\\
C_l &=&C_1 +\frac{{\bf q}^2}{q^4}C_2\, .
\label{clt}
\end{eqnarray}
At $T=0$ these two functions are not independent, $C_{l}=C_{t}/q^2=C_{V}$.
They are also related at $T\neq 0$, if \\
{}~\\
$\underline{{\bf q}=0,\; q_0 \neq 0,\; Q^2=-q_0^2}$

\begin{eqnarray}
\lefteqn{C_{1}^{V}(T)-C_{1}^{V}(T=0)=C_l^T -C_l^{T=0}=
-\frac{1}{Q^2}\left( C_t^T-C_t^{T=0}\right)} \nonumber\\
& & =-\left(\epsilon -\frac{1}{2}\epsilon^2\right)
(C_{1}^{V}-C_{1}^{A})+c\,\frac{T^4}{2Q^2},
\label{q0}
\end{eqnarray}
where $c$ was defined in Eq.~(\ref{coef}).

In another special case \\
{}~\\
$\underline{q_0 =0,\; {\bf q}\neq 0,\; Q^2={\bf q}^2}$

\begin{eqnarray}
Q^2 (C_l^T-C_l^{T=0})&=&
{}~~\left(\epsilon -\frac{1}{2}\epsilon^2\right) (C_{V}-C_{A})+
c\,\frac{T^4}{2Q^2} \nonumber\\
C_t^T-C_t^{T=0}&=& -\left(\epsilon -\frac{1}{2}\epsilon^2\right) (C_{V}-C_{A})
+c\,\frac{T^4}{2Q^2}\, ,
\label{qvec}
\end{eqnarray}

%ins8
Till now we considered the correlators at negative $q^2 =-Q^2$. In order to
interpret the results in terms of the particle mass shifts, the amplitudes
at negative $q^2$ must be represented using dispersion relations through the
contributions of physical states, defined at positive $q^2$ and $s$. Unlike
the case of $T=0$, where the correlators are functions of one variable,
$q^2$, at $T\neq 0$ they are functions of two variables, $q_0$ and
${\bf q}^2$. In this case the only way to represent the amplitude at
negative $q^2$ through the contributions of physical states is to use the
dispersion relation in $q_0$ at fixed ${\bf q}^2$. (In the opposite case,
when $q_0$ is fixed and ${\bf q}^2$ is variable, the amplitudes would have
non-physical singularities).

So, let us consider the case ${\bf q}=0$, $q_0\neq 0$, $Q^2 =-q_{0}^2$.
In this case there is only one structure function,
$C_{l} (q_0)=C_1 (q_0)$. We choose the standard model for the spectral
densities as a sum of the lowest resonances and continuum. The dispersion
relations over $q_0$ are contributed by the physical states in the $q^2$-
and $s$-channels. Therefore, for the structure function $C_{1}^{V}(Q^2,T)$
we have

\begin{equation}
C_{1}^{V}(Q^2,T)=\frac{\lambda_{\rho ,V}^{2}(T)}{Q^2+m_{\rho}^{2}(T)}+
\frac{\lambda_{a_{1},V}^{2}(T)}{Q^2+m_{a_{1}}^{2}(T)}+
\frac{1}{\pi}\int_{s_0}^{\infty}\frac{\rho_{V}(s,T)}{Q^2+s}ds,
\label{c1v}
\end{equation}
where the first and the second terms in the r.h.s. of Eq.~(\ref{c1v})
correspond to the contributions of $\rho$ and $a_1$ mesons to the vector
current correlator, $\lambda_{\rho ,V}^{2}(T)$ and $\lambda_{a_1 ,V}^{2}(T)$
are the corresponding $T$-dependent coupling constants, and
$\lambda_{a_1, V}\sim T^2$. (The subtraction constant is omitted). A
similar equation holds for the axial current correlator $C_{1}^{A}(Q^2 ,T)$.
{}From Eq.~(\ref{q0}) it is easy to see that the terms of order $T^4/Q^2$
vanish in the difference $C_{1}^{V}(Q^2 ,T)-C_{1}^{A}(Q^2 ,T)$:

\begin{eqnarray}
\lefteqn{C_{1}^{V}(Q^2 ,T)-C_{1}^{A}(Q^2 ,T)-
\left[C_{1}^{V}(Q^2 ,0)-C_{1}^{A}(Q^2 ,0)\right]} \nonumber \\
& & =-\epsilon\left( 1-\frac{\epsilon}{2}\right)
\left[ C_{1}^{V}(Q^2 ,0)-C_{1}^{A}(Q^2 ,0)\right]\, .
\label{32}
\end{eqnarray}
If we put

\begin{equation}
m_{\rho}^{2}(T)=m_{\rho}^{2}+\delta m_{\rho}^{2}, ~~~
m_{a_1}^{2}(T)=m_{a_1}^{2}+\delta m_{a_1}^{2},
\label{33}
\end{equation}
then from Eqs.~(\ref{c1v}) and (\ref{32}) it follows that

\begin{equation}
\lambda_{\rho}^{2}\delta m_{\rho}^2 -\lambda_{a_1}^{2}\delta m_{a_1}^2 =0
\label{34}
\end{equation}
and

\begin{eqnarray}
\frac{\lambda_{\rho ,V}^{2}(T)}{\lambda_{\rho}^2}=
\frac{\lambda_{a_1 ,A}^{2}(T)}{\lambda_{a_1}^2}&=&
1-\epsilon\left( 1-\frac{1}{2}\epsilon\right) \nonumber \\
\frac{\lambda_{\rho ,A}^{2}(T)}{\lambda_{\rho}^2}=
\frac{\lambda_{a_1 ,V}^{2}(T)}{\lambda_{a_1}^2}&=&
\epsilon\left( 1-\frac{1}{2}\epsilon\right)\, .
\label{35}
\end{eqnarray}
For the sum $C_{1}^{V}(Q^2 ,T)+C_{1}^{A}(Q^2 ,T)$ from Eq.~(\ref{q0}) we
have

\begin{equation}
C_{1}^{V}(Q^2 ,T)+C_{1}^{A}(Q^2 ,T)-
\left[C_{1}^{V}(Q^2 ,0)+C_{1}^{A}(Q^2 ,0)\right]=c\,\frac{T^4}{Q^2}\, .
\label{36}
\end{equation}
It is clear from the comparison of Eqs.~(\ref{c1v}) and (\ref{36}) that with
our model of hadronic spectrum the continuum cannot contribute to the l.h.s.
of Eq.~(\ref{36}), since the imaginary part of the r.h.s. vanishes at
$Q^2 <0$. Then from Eqs.~(\ref{c1v}) and (\ref{36}) we have

\begin{equation}
\lambda_{\rho}^{2}\delta m_{\rho}^2 +\lambda_{a_1}^{2}\delta m_{a_1}^2 =
-c\, T^4\, .
\label{37}
\end{equation}
(Taking into account only terms $\sim T^4$, we put
$\lambda_{\rho}^2 (T)\approx\lambda_{\rho}^2 (0)=\lambda_{\rho}^2$ and
$\lambda_{a_1}^2 (T)\approx\lambda_{a_1}^2 (0)=\lambda_{a_1}^2$). From
Eqs.~(\ref{34}) and (\ref{37}) we have

\begin{equation}
\delta m_{\rho}^2 = -c\,\frac{T^4}{2\lambda_{\rho}^2}, ~~~
\delta m_{a_1}^2 = -c\,\frac{T^4}{2\lambda_{a_1}^2}\, .
\label{38}
\end{equation}
The residues in Eq.~(\ref{38}) are the standard couplings of $\rho$ and
$a_1$ mesons with the vector and axial currents,
$\lambda_{\rho}^2 =m_{\rho}^2/g_{\rho}^2$,
$\lambda_{a_1}^2 =m_{a_1}^2/g_{a_1}^2$. Numerically they are rather close,
$\lambda_{\rho}^2\approx\lambda_{a_1}^2\approx 0.02$~\cite{dei}.

We see that both the $\rho$ and $a_1$ masses start decreasing with $T$, and
the mass shifts appear in order $T^4$ and, in terms of OPE, are due to
Lorentz non-scalar condensates as emphasized in refs.\cite{eek,e4}
\footnote{In Ref.~\cite{hkl} it was claimed that the $\rho$ and $a_1$
masses move in order $T^2$. This conclusion comes from an incorrect choice
of the spectral density in the sum rules: the mixing of vector and axial
channels was not taken into account. For this reason the results obtained
in this paper for the $T$-dependence of the masses are not reliable,
though we agree with the analysis of $T$-dependent condensates carried out
there.}.
The corrections proportional to powers of $T^2/F_{\pi}^2$ affect
only the residues of the currents. This fact can be easily
understood. Indeed, in the OPE for the correlators taking into account
finite temperatures to order $T^2$ would result only in the change of the
same Lorentz scalar condensates which appear in OPE at $T=0$.  Then it is
clear from the representation through dispersion relations that any such
change can be described by modifications of the residues without affecting
the position of poles.

The scenario when both the vector and axial masses
decrease with $T$ is allowed by Weinberg sum rules at $T\neq 0$\cite{ks}.
Numerically the mass shifts are rather small. Even at $T=150--200\,$MeV
(usually accepted values for the phase transition temperature)
$\delta m_{\rho}^2\approx\delta m_{a_1}^2\sim -(0.01~-~0.02)\,$GeV$^2$. At
the same time at these temperatures the change of residues according to
Eq.~(\ref{35}) is very essential. (For this reason Eqs.~(\ref{38}) are not
completely reliable at $T>100\,$MeV, since we put
$\lambda^2 (T)\approx\lambda^2 (0)$).

Let us summarize our main results. The corrections of order $T^4$ to the
correlators of vector and axial currents were calculated in QCD in a model
independent way in the chiral approximation when $u$, $d$ quarks and pions
are massless. The results are expressed in terms of the second moment of the
pion structure function in deep inelastic lepton-pion scattering that is
equal to the matrix element of the quark energy-momentum tensor over the
pion state, or to the fraction of the pion momentum carried by quarks in the
parton model. Interpreted in terms of physical mesons the calculated
corrections correspond to negative mass shifts of the $\rho$ and $a_1$
mesons proportional to $T^4$. (As was shown earlier~\cite{ei,dei}, the mass
shifts are absent in order $T^2$). These mass shift originate from Lorentz
non-scalar condensates in OPE. Numerically they are rather small at
$T\leq 100\,$MeV, where our approach is correct. The corrections arising
from a finite pion mass were not touched in this paper. We plan to consider
them in a future work.

We gratefully acknowledge useful discussions with H. Leutwyler.
This work was supported in part by the International Science Foundation
under Grant No. M9H000. The work of V.L.E. was supported in part by
Schweizerischer Nationalfonds.

\end{document}